\documentclass[%
 reprint,
 amsmath,amssymb,
 aps,
]{revtex4-2}

\usepackage{graphicx}
\usepackage{dcolumn}
\usepackage{bm}

\usepackage{amssymb}
\usepackage{lipsum}
\usepackage{soul}
\usepackage{slashed}
\usepackage{amsmath}
\usepackage{bbold}
\usepackage[compat=1.0.0]{tikz-feynman}
\usetikzlibrary{decorations.markings,decorations.pathreplacing}

\tikzfeynmanset{
	fermion2/.style={
		/tikz/postaction={
			/tikz/decorate=false,
		},
	},
}
\tikzfeynmanset{
	fermion3/.style={
		/tikz/decoration={
			markings,
			mark=at position 0.5 with {
				\node[
				dot,
				fill,
				draw=none,
				] { };
			},
		},
		/tikz/postaction={
			/tikz/decorate=true,
		},
	},
}

\begin{document}

\preprint{APS/123-QED}

\title{The Quark Structure of the Nucleon and Moat Regimes in Nuclear Matter}

\author{Theo F. Motta}
\email{theo.motta@unesp.br}
\author{Gastão Krein}
\email{gastao.krein@unesp.br}
\affiliation{Instituto de Física Teórica, Universidade Estadual Paulista, Rua Dr. Bento Teobaldo Ferraz, 271 - Bloco II - 01140-070 São Paulo, SP, Brazil}

\date{\today}

\begin{abstract}
We employ a model of nuclear structure that takes into account the quark substructure of the baryons to understand the behavior of static two-point correlation functions of meson fields in dense nuclear matter. We show that these correlation functions display ``moat'' regimes, where the inverse static correlation function is nonmonotonic in momentum space. This can have interesting consequences for the nature of nuclear matter, and it strengthens the possibility of a liquid-crystal phase for high densities. 
\end{abstract}

\maketitle

\section{Introduction}
\label{introduction}
Already decades ago, model descriptions of nuclear matter at saturation density had reached a point where the bulk properties of nuclei (e.g. binding energy, density profile, etc.) could be adequately explained. Naturally, calculating nuclear properties with high precision across a large range of Z and A requires sophisticated models, which are still under investigation today. Nevertheless, by all accounts, infinite symmetric nuclear matter shows a minimum of binding energy per nucleon for a finite baryon number density, the so-called saturation density~$n_0$, causing the density at the core of {large} nuclei to saturate around $n_0$. For larger densities, things tend to get more and more uncertain. Variations of the so-called Relativistic Mean-Field (RMF) models {\cite{Serot:1997xg}} have been used to study high-density matter, which is expected to be realized in the core of neutron stars. However, since the parameters of the models are fixed at low densities, the extrapolation is uncertain. Density-dependent couplings and the in-medium properties of baryons that only appear at high densities (hyperons, for instance) cannot be easily determined within RMF models.

Going beyond mean-field descriptions of infinite nuclear matter can be even more nontrivial. Take, for example, Relativistic Random Phase Approximation (RRPA) approaches, where the meson propagators are dressed to {one}-loop order. RMF models show zeros of the inverse meson propagators for vanishing frequency and finite three-momentum, for densities of about $\approx 0.4$\,fm$^{-3}$. In principle, this would indicate that homogeneous infinite nuclear matter is unstable with respect to inhomogeneous fluctuations of the density and that the true ground state is crystalline. This issue is well documented in the literature, for example, the authors of the review article in Ref.~\cite{Serot:1991st} remark that to really see if this instability is real, among other things, ``\textit{the composite structure of the baryon must be taken into account}''. 

One well established model of nuclear interactions that does take into account the quark substructure of the nucleon is the Quark-Meson Coupling (QMC) model. It provides an interesting microscopic mechanism for saturation \cite{Guichon:1987jp,Guichon:1995ue}, gives excellent agreement with experimental data on heavy nuclei \cite{Rikovska-Stone:2006gml} and adequately explains neutron stars with heavy masses without falling into the so-called ``hyperon puzzle'' \cite{Motta:2019tjc,Leong:2023yma}. The QMC model contains density-dependent interactions that can be intuitively explained without the need for new coupling parameters. In this paper, we show that the model also seems to help shed some light on this issue with the stability of nuclear matter in RRPA. Furthermore, in every parameter space region where the instability is not present, it appears to show that dense nuclear matter manifests a strongly pronounced moat regime \cite{Pisarski:2021qof}, where the static correlation functions of density operators (number density, static density, etc.) show some oscillatory behavior in configuration space -- non-monotonic behavior in momentum space. There is good evidence that some type of moat-regime takes place in dense quark-matter \cite{Fu:2019hdw,cao2025moatregimes21flavor,Fu_2025,töpfel2024phasestructurequarkmatter} but so far, it has not been looked for in models of nuclear matter.
In the next sections, we will first briefly explain the QMC model, then define and discuss the implications of the so-called moat regime, and show results for RRPA.

\section{The QMC Model}

The quark-meson coupling model describes a system of many nucleons as a collection of MIT-bag-like objects where the quarks inside different baryons interact with each other by exchanging fundamental meson fields. 
It is important to stress, though, that any confining model can be used in order to build a QMC-like model. In the review article \cite{Motta:2022nlj} these variations were collected and discussed in detail. Nevertheless, the phenomenology is similar and we show here the classical version with the MIT bag model.
In the mean-field approximation, one finds that the equation of motion for a quark inside the bag is simply:
\begin{equation}\label{diraceq}
    (i\slashed \partial -m_q + g_\sigma^q\sigma - g_\omega^q\gamma^0\omega_0)\psi=0, 
\end{equation}
subject to the spherical boundary conditions
\begin{equation}\label{bagbondary}
    i\gamma\cdot{\hat r}\psi(R)=-\psi(R),
\end{equation}
where R is the bag radius and $\hat r$ the vector normal to the surface of the sphere. Equation (\ref{bagbondary}) simply means that the fermion is reflected elastically from the bag surface. The lowest energy time-independent solution of Eq.~(\ref{diraceq}), i.e., the wave function of the quark inside the hadron, is
\begin{equation}
\psi_m(\vec{r})=\mathcal{N}\,\binom{j_0\left(x r / R\right)}{i \beta_q \vec{\sigma} \cdot \hat{r} j_1\left(x r / R\right)} \frac{\chi_m}{\sqrt{4 \pi}},
\end{equation}
where $\chi_m$ is a spinor that describes the inner degrees of freedom of the quark (spin, flavor, and color), $j_0$ and $j_1$ are spherical bessel functions. The value of $x$ satisfies the boundary conditions
\begin{equation}
    j_0(x)=\beta_q j_1(x).
\end{equation}
where $\beta_q$ is
\begin{equation}
\begin{aligned}
\beta_q=\sqrt{\frac{\Omega-m_q R}{\Omega+m_q R}},\quad \Omega & =\sqrt{x^2+\left(m_q R\right)^2},
\end{aligned}
\end{equation}
and the wave function is normalized by
\begin{equation}
    \mathcal{N}^{-2}  =2 R^3 j_0^2(x)\left[\Omega(\Omega-1)+m_q R / 2\right] / x^2 .
\end{equation}
One can easily calculate the energy of a bag to be simply
\begin{equation}
M=\frac{N_u \Omega_u+N_d \Omega_d+N_s \Omega_s}{R}+ \mathcal{B} V.
\end{equation}
where $\mathcal{B}$ is the bag pressure, $V$ its volume, and $N_q$ the number of quarks of each flavor. Note that the energy eigenvalues of the quarks $\Omega_q/R$ are now dependent on the sigma mean-field. This dependence turns out non-linear and can be interpreted as emergent many-body forces between the baryons. Of course, in order for the model to accurately resolve the mass of the full baryon decuplet, more detailed hadronic structure effects have to be taken into account. These include, for instance, fluctuations of the gluon field
corrected for by a zero-point energy parameter, and also one-gluon exchanges between the quarks to correct for the spin-induced mass splitting. For detailed reviews of the model, the reader is referred to Refs.~\cite{Motta:2022nlj,Guichon:2018uew}. For the purposes of this letter, suffice to understand that by taking into account the inner structure of the baryon, the QMC model predicts a non-linear behavior of the baryon mass as a function of the scalar mean-field. Ultimately, a quadratic function approximates excellently the mass as a function of the sigma field
\begin{equation}\label{Mstar}
    M_B^\star(\sigma) = M_B - g_{\sigma }\sigma + \frac{d_{\sigma }}{2}(g_{\sigma }\sigma)^2.
\end{equation}
The coupling constant $g_{\sigma }$ is fixed to fit saturation properties and $d_{\sigma }$ is the scalar polarizability, which is calculated, not a free parameter. It is a direct consequence of the quark inner structure of the baryon. Both $g_{\sigma }$ and $d_{\sigma }$ can in principle be different for each baryon of the decuplet. For the nucleon, as a function of the bag radius, the scalar polarizability is well approximated by
\begin{equation}\label{dsigma}
    \frac{d_\sigma}{2}=0.0022\,\text{fm}+0.1055 \,R_N-0.0178\text{\,fm}^{-1}\,{R_N^2}.
\end{equation}

The QMC model has been quite successful and provides a very clear picture of nuclear many-body physics. With very few parameters (usually around 5, depending on the many variations of the model), one can find unreasonably accurate values for the binding energy of heavy nuclei \cite{Stone:2016qmi}. This indicates that, although rather simple, the model is able to capture complex dynamics of many-nucleon systems. It is also successful in describing neutron stars cores \cite{Motta:2019tjc,Rikovska-Stone:2006gml} and crust \cite{Antic:2020zuk,Kalaitzis:2019dqc}. 

In the following sections, we address a new issue that has been growing in interest by the community, namely, the spatial configuration of the correlation functions. When the occupation number density or the scalar density of many fermion systems is space-dependent, the medium is in an inhomogeneous or crystalline phase. If the two-point function of the scalar density ($\bar\psi \psi$) or number density operators ($\bar\psi\gamma_0\psi$) is spatially dependent, i.e.
$$
\begin{aligned}
    \langle \bar\psi(x)\gamma_0\psi(x)\bar\psi(y)\gamma_0\psi(y)\rangle =& s(x-y)
\\
\langle \bar\psi(x)\psi(x)\bar\psi(y)\psi(y)\rangle =& f(x-y),
\end{aligned}
$$
where $s$ and $f$ are generic functions of $x-y$,
the system is in the so-called ``moat-regime'' \cite{Rennecke:2021ovl,Pisarski:2018bct,Pisarski:2021qof,Pannullo:2024hqj,Rennecke:2023xhc}. The name comes from the fact that the inverse of these functions, in momentum space, resembles a moat, manifesting a nonzero minimum at the effective wave number of the modulated $s$ and $f$ functions. Although every inhomogeneous phase will manifest a moat regime, the latter is more general.

\section{Inhomogeneous Phases and the Moat Regime}
As briefly mentioned above, one might define an inhomogeneous phase \cite{Buballa:2014tba}, generally, as one where the expectation value of the fermion field bilinears becomes space dependent, i.e. the scalar density,  pseudo-scalar density, the number density, etc.
\begin{equation}
    \begin{aligned}
        \langle\bar\psi(\vec x)\psi(\vec x)\rangle&=n_s(\vec x), \\
        \langle\bar\psi(\vec x)i\gamma_5\tau_3\psi(\vec x)\rangle&=n_{ps}(\vec x), \\
        \langle\bar\psi(\vec x)\gamma_0\psi(\vec x)\rangle&=n(\vec x),\,\dots
    \end{aligned}
\end{equation}
In nuclear matter this has been thought of by Migdal in the 70's \cite{Migdal:1974gdq} in the context of $\pi$-condensation. Lately, however, much attention has been paid to the possibility of inhomogeneous chiral phases in quark matter \cite{Buballa:2014tba,Pannullo:2024hqj,Ferrer:2021mpq}. Regardless of the nature of matter, interacting fermionic systems can develop inhomogeneous phases at high densities. The standard way to look for these is by means of a stability analysis, i.e., testing the stability of the homogeneous ground state against small inhomogeneous fluctuations. In the context of simplified models that can be bosonised, this simply entails writing down the free energy as a functional of the mean bosonised fields $\Omega[\phi( x)]$ and expanding about the homogeneous stationary point $\bar\phi$. Explicitly, we expand about $\phi( x) \rightarrow \bar\phi + \delta\phi(\vec x)$ and calculate
\begin{equation}
    \Omega[\phi( x)] \approx \Omega[\bar\phi] + \delta\Omega[\bar\phi,\delta\phi(\vec x)].
\end{equation}
If $\delta \Omega$ is positive, the free energy grows and the homogeneous system is stable. If it is negative, however, the homogeneous system is unstable with respect to the formation of inhomogeneous structures. In mean-field models, one inevitably obtains $\delta\Omega$ to be
\begin{equation}\label{eq:mfSA}
    \delta\Omega[\bar\phi,\delta\phi(\vec x)]=
    \int_{\vec{q}} \vert \delta\phi(\vec q) \vert^2 \, D_\phi^{-1}(q_0=0,\vec{q}).
\end{equation}
To break it down, Eq.~(\ref{eq:mfSA}) simply involves an integral over the three-momentum of the modulus squared of the $\delta\phi$ in momentum space (which is always positive!), times the inverse propagator of the bosonised field at zero frequency and finite wave-vector $\vec q$. Clearly, then, since the mod-squared of $\delta\phi$ is always positive, whenever there is a region of negativity of $D_\phi^{-1}(0,\vec q)$ for any finite three-momentum, we know that the system is unstable. Note that in order to conclude this, one does not need to specify $\delta\phi(\vec x)$ explicitly.
For more complicated models, there are generalizations of the mean-field stability analysis \cite{Motta:2023pks,Motta:2024agi} which can also be applied to truncations of QCD in order to investigate inhomogeneous quark phases \cite{Motta:2024rvk}.

In summary, when the static inverse meson propagator is negative only for finite three-momentum, an instability towards an inhomogeneous phase is certainly present. More generally, though, when the static inverse meson propagator is \textit{nonmonotonic}, this configures a moat regime. In such a scenario the expectation value of the bosonised field $\langle \phi(\vec x) \rangle$ could be constant, but the two-point function $\langle \phi(0) \phi(\vec x) \rangle$ is an oscillating function on $\vec x$ which decays for large $\vec x$.

The main point of this paper is to perform this exact analysis. We calculate the dressed inverse meson propagator for symmetric nuclear matter, within the QMC model, including scalar ($\sigma$) and vector ($\omega$) meson exchanges.

\section{RRPA}
At zero temperature, in a dense medium, the mean-field dressed propagator of the nucleons is given by
\begin{equation}
    \label{fermprop2}
    \begin{aligned}
        G(k) =& \Bigg[
    \frac{\slashed k^\star + M^\star}{k^{\star\,2} - M^{\star\,2} +i\epsilon}
    +
    \Bigg(\frac{\slashed k^\star + M^\star}{k^{\star\,2} - M^{\star\,2} -i\epsilon}
    \\&
    \,\,-\frac{\slashed k^\star + M^\star}{k^{\star\,2} - M^{\star\,2} +i\epsilon}
    \Bigg)\theta(k_0^\star)\theta(k_F - |\vec{k}|)
    \Bigg],
    \end{aligned}
\end{equation}
where $M^\star$ is the effective nucleon mass (which for us will be given by Eq.~(\ref{Mstar})), $k_\mu^\star=(k_0-g_\omega \omega,\,\vec k)$ and $k_F$ defines the Fermi momentum which for symmetric nuclear matter is written as
\begin{equation}
    k_F=\left(\frac{3\pi^2 n}{2}\right)^{\frac{1}{3}}.
\end{equation}
Random phase approximation basically entails dressing the bosonic propagators as well, to 1-loop order, i.e., solving the following equation
\begin{equation}\label{eq:dse}
    \begin{tikzpicture}
        \begin{feynman}
            \vertex (a);
            \vertex [right=0.5cm of a] (ap);
            \vertex [right=1cm of a] (b);
            \diagram*{
                (a) -- [boson] (b);
            };
            \draw (ap) node [dot];
        \end{feynman}
    \end{tikzpicture}^{-1}
    \,\,=\,\,
    \begin{tikzpicture}
        \begin{feynman}
            \vertex (a);
            \vertex [right=0.5cm of a] (ap);
            \vertex [right=1cm of a] (b);
            \diagram*{
                (a) -- [boson] (b);
            };
        \end{feynman}
    \end{tikzpicture}^{-1}
    \,\,-\,\,
    \raisebox{-0.27cm}{
    \begin{tikzpicture}
    \begin{feynman}
    \vertex (a);
    \vertex [right=0.25cm of a] (ap);
    \vertex [right=0.5cm of ap] (bp);
    \vertex [right=0.25cm of bp] (b);
    \vertex [above= 0.2cm of ap] (up);
    \vertex [below= 0.2cm of ap] (down);
    \vertex [right=0.25cm of up] (t1);
    \vertex [right=0.25cm of down] (t2);
    \diagram*{
    	(a) --  [boson] (ap);
    	(ap) -- [fermion2, half left] (bp);
    	(ap) -- [fermion2, half right] (bp);
    	(bp) -- [boson] (b);
    };
    \draw (t1) node [dot];
    \draw (t2) node [dot];
    \end{feynman}
    \end{tikzpicture}}
\end{equation}
where the dotted fermion lines are given by Eq.~(\ref{fermprop2}) and the dotted wiggly lines denote the dressed meson propagators.
In principle, since we are considering both $\sigma$ and $\omega$ exchanges at finite density, these two channels will mix, so the free bosonic propagator is best written as a 5x5 matrix
\begin{equation}
    \begin{tikzpicture}
        \begin{feynman}
            \vertex (a);
            \vertex [right=0.5cm of a] (ap);
            \vertex [right=1cm of a] (b);
            \diagram*{
                (a) -- [boson] (b);
            };
        \end{feynman}
    \end{tikzpicture}=
    \left(\begin{matrix}
        \Delta^0 & 0 \\
        0 & D^0_{\mu\nu}
    \end{matrix}\right),
\end{equation}
where
\begin{equation}
    \begin{aligned}
        &\Delta^0(q)=\frac{1}{q^2-m_\sigma^2},
    \quad
    D^0_{\mu\nu}(q) = {\left(\eta_{\mu\nu}-\frac{q_\mu q_\nu}{q^2}\right)}
    D^0(q),\\
    &D^0(q)=\frac{-1}{q^2-m_\omega^2},
    \end{aligned}
\end{equation}
and the self-energy contains off-diagonal terms (for the elements of this matrix, and throughout this paper, unless explicitly stated, we use the same conventions as in Ref.~\cite{Lim:1989ma})
\begin{equation}
    \raisebox{-0.25cm}{
    \begin{tikzpicture}
    \begin{feynman}
    \vertex (a);
    \vertex [right=0.25cm of a] (ap);
    \vertex [right=0.5cm of ap] (bp);
    \vertex [right=0.25cm of bp] (b);
    \vertex [above= 0.2cm of ap] (up);
    \vertex [below= 0.2cm of ap] (down);
    \vertex [right=0.25cm of up] (t1);
    \vertex [right=0.25cm of down] (t2);
    \diagram*{
    	(a) --  [boson] (ap);
    	(ap) -- [fermion2, half left] (bp);
    	(ap) -- [fermion2, half right] (bp);
    	(bp) -- [boson] (b);
    };
    \draw (t1) node [dot];
    \draw (t2) node [dot];
    \end{feynman}
    \end{tikzpicture}
    }=
    \left(\begin{matrix}
        \Pi_s & \Pi_{\mu\phantom{\mu}} \\
        \Pi_{\nu} & \Pi_{\mu\nu}
    \end{matrix}\right).
\end{equation}

Note that if the channels did not mix, the dressed boson propagator would be diagonal. The moat-regimes and instabilities could be extracted simply by looking at each of the eigenvalues (i.e. propagators for each mesonic channel) individually. This not being the case, the fully dressed propagator has to be diagonalised. However, an equivalent and more direct way to extract the instabilities is by means of the dielectric function, which we define in the following section.

\section{The Static Dielectric Function}
The equation for the boson propagator, Eq.~(\ref{eq:dse}), can be written equivalently as:
\begin{equation}\label{eq:dse2}
    \begin{tikzpicture}
        \begin{feynman}
            \vertex (a);
            \vertex [right=0.5cm of a] (ap);
            \vertex [right=1cm of a] (b);
            \diagram*{
                (a) -- [boson] (b);
            };
            \draw (ap) node [dot];
        \end{feynman}
    \end{tikzpicture}
    =
    \frac{\hfil\raisebox{0.1cm}{\begin{tikzpicture}
        \begin{feynman}
            \vertex (a);
            \vertex [right=0.5cm of a] (ap);
            \vertex [right=1cm of a] (b);
            \diagram*{
                (a) -- [boson] (b);
            };
        \end{feynman}
    \end{tikzpicture}}\hfil}{
    \mathbb{1}-
    \raisebox{-0.22cm}{
    \begin{tikzpicture}
    \begin{feynman}
    \vertex (a);
    \vertex [right=0.90cm of a] (ap);
    \vertex [right=0.5cm of ap] (bp);
    \vertex [right=0.25cm of bp] (b);
    \vertex [above= 0.2cm of ap] (up);
    \vertex [below= 0.2cm of ap] (down);
    \vertex [right=0.25cm of up] (t1);
    \vertex [right=0.25cm of down] (t2);
    \diagram*{
    	(a) --  [boson] (ap);
    	(ap) -- [fermion2, half left] (bp);
    	(ap) -- [fermion2, half right] (bp);
    	(bp) -- [boson] (b);
    };
    \draw (t1) node [dot];
    \draw (t2) node [dot];
    \end{feynman}
    \end{tikzpicture}}},
\end{equation}
which is only nonsingular if the determinant of the denominator is finite nonzero. Therefore, a necessary and sufficient condition for the appearance of poles in the meson propagator is that the so-called dielectric function:
\begin{equation}
    \epsilon(q_0,\vec{q})=
    \text{det}\left(
    \mathbb{1}-
    \raisebox{-0.22cm}{
    \begin{tikzpicture}
    \begin{feynman}
    \vertex (a);
    \vertex [right=0.90cm of a] (ap);
    \vertex [right=0.5cm of ap] (bp);
    \vertex [right=0.25cm of bp] (b);
    \vertex [above= 0.2cm of ap] (up);
    \vertex [below= 0.2cm of ap] (down);
    \vertex [right=0.25cm of up] (t1);
    \vertex [right=0.25cm of down] (t2);
    \diagram*{
    	(a) --  [boson] (ap);
    	(ap) -- [fermion2, half left] (bp);
    	(ap) -- [fermion2, half right] (bp);
    	(bp) -- [boson] (b);
    };
    \draw (t1) node [dot];
    \draw (t2) node [dot];
    \end{feynman}
    \end{tikzpicture}}
    \right),
\end{equation}
is zero. For our intents and purposes, it is necessary only to look at the Static Dielectric Function (SDF) $\epsilon(0,\vec{q})$, which for simplicity we will call $\epsilon(\vec{q})$. Since the free-meson propagator is both diagonal and monotonic, we can also use the SDF to identify moat regimes. Whenever the SDF is not monotonic, some eigenvalue (or eigenvalues) of the diagonalised dressed propagator matrix must be non-monotonic.

Finally, it is useful to separate the SDF into longitudinal and transverse parts in order to identify which mesonic component is causing the instabilities, they are defined respectively as (omitting the momentum arguments)
\begin{equation}
\begin{array}{l}
\epsilon_{\mathrm{L}} \equiv\left(1-\Delta^0 \Pi_{\mathrm{s}}\right)\left(1-D^0 \Pi_{\mathrm{L}}\right)+\frac{q_\mu^2}{q^2} \Delta^0 D^0\left(\Pi_0\right)^2, \\
\epsilon_{\mathrm{T}} \equiv\left(1+D^0 \Pi_{\mathrm{T}}\right),
\end{array}
\end{equation}
where $\Pi_L = \Pi_{00} - \Pi_{11}$ and $\Pi_T=\Pi_{22\text{ or }33}$ (see also \cite{Lim:1989ma,Saito:1998wd}). The full dielectric function is $\epsilon=\epsilon_L\epsilon_T^2$, however, the zeros of $\epsilon_{L,T}$ individually are sufficient for $\epsilon$ to be zero; therefore, we can plot them independently.

\section{Results}
We chose to present our results with two different parametrizations of the QMC model, one with a heavier $\sigma$ mass of $700$~MeV and one with a lighter $600$~MeV scalar channel, which is the value usually taken in the model to study neutron star matter. We believe these two parameter sets are enough to see the important trends. We use a scalar polarizability of $d_\sigma=0.15$~fm, which, according to Eq.~(\ref{dsigma}), obtained from Ref.~\cite{Rikovska-Stone:2006gml}, corresponds to a bag radius of $0.8$fm. The parameters were fitted to reproduce a saturation density of $0.16$fm$^{-3}$ and a binding energy of $-15.9$MeV.

\subsection{Instabilities}
\begin{figure}
    \centering
    \includegraphics[width=\linewidth]{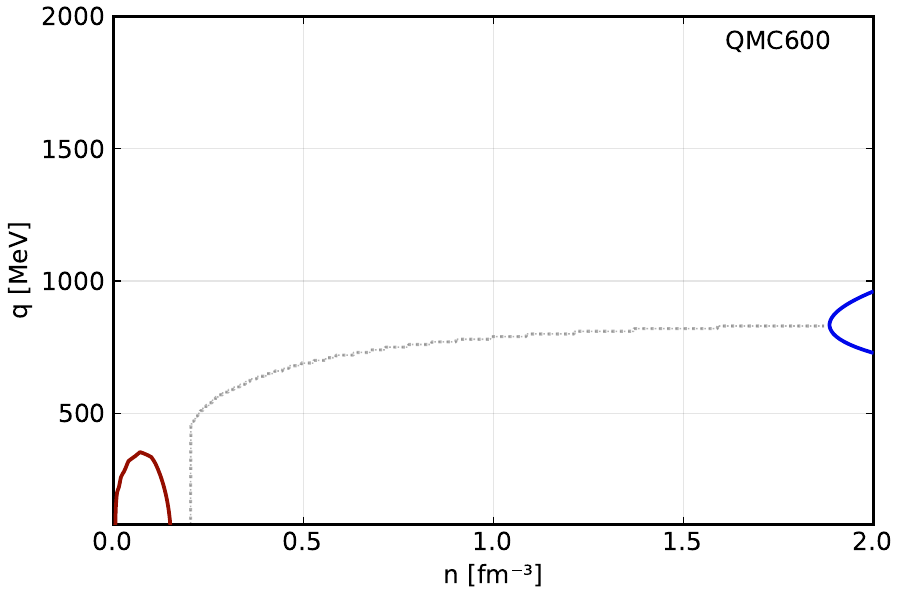}
    \includegraphics[width=\linewidth]{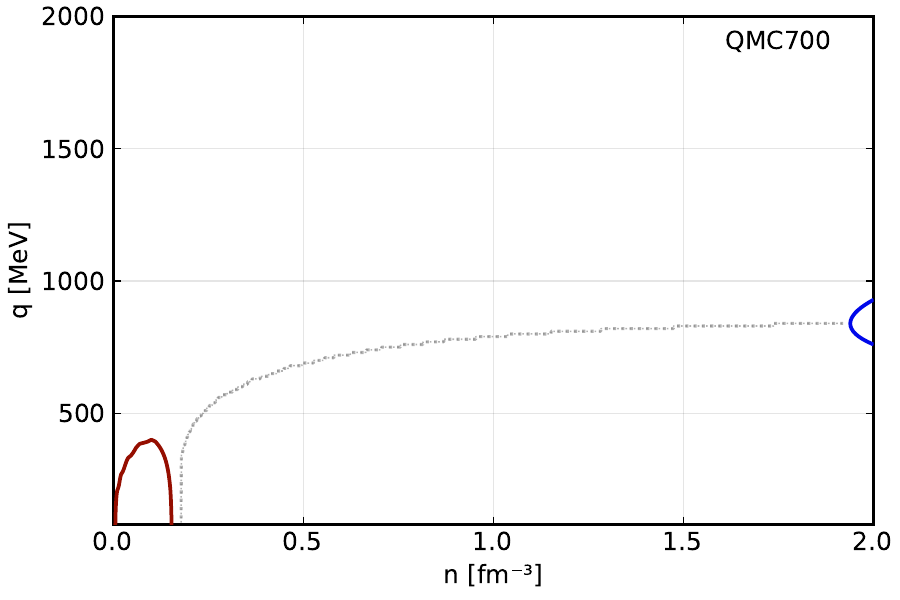}
    \caption{Zeros of the SDF as a function of the momentum and density for symmetric nuclear matter. The zeros of the transverse part is shown in blue and the longitudinal zeros are shown in dark-red.}
    \label{fig:qmc_sym}
\end{figure}

In Fig.~\ref{fig:qmc_sym} we show the zeros of the SDF, i.e., the instability regions, for two parametrizations of the QMC model defined in Table~\ref{tab:params}. The blue lines are the zeros of the longitudinal part of the SDF and the dark-red lines are the zeros of the transverse part.

\begin{table}
    \centering
    \begin{tabular}{|c|ccc|}\hline
        \, & $g_\sigma$ & $g_\omega$ & $m_\sigma$ \\ \hline \hline
        QMC600 & 9.02 & 8.42 & 600~MeV\\
        QMC700 & 10.52 & 8.41 & 700~MeV\\\hline
    \end{tabular}
    \caption{Parameter sets for the QMC model. For both sets the mass of the $\omega$ channel is kept at $783$~MeV.}
    \label{tab:params}
\end{table}
\begin{figure}
    \centering
    \includegraphics[width=\linewidth]{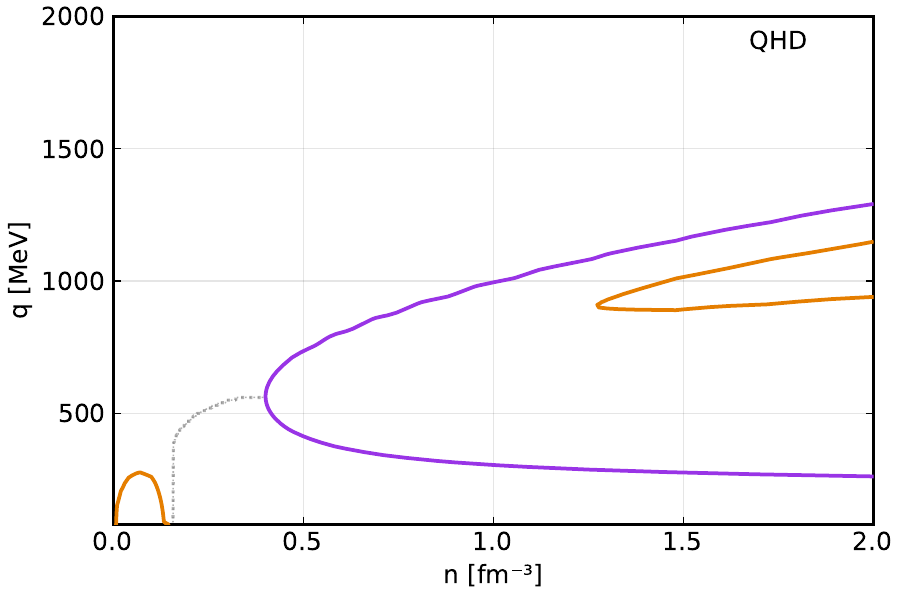}
    \caption{Zeros of the SDF, analogous to Fig.~\ref{fig:qmc_sym} for QHD, reproducing Ref.~\cite{Lim:1989ma}. Here in purple we have the transverse SDF and in orange, the longitudinal SDF.}
    \label{fig:qhd}
\end{figure}

Note that for lower densities we have a small region of instability ending just before the saturation density. This corresponds to the region around the liquid-gas transition where the system forms droplets of denser matter on top of a lower-density gas. The most interesting result comes at higher densities. We find that the longitudinal instabilities vanish completely.
Only at unphysically high densities can we see the onset of transverse instabilities.
According to the framework of the stability analysis laid out above, this would indicate that the system is unstable with respect to inhomogeneous perturbations of the number density and scalar density. Of course, at such high densities, the model is surely unserviceable and full QCD must be employed.

The standard RMF result is shown in Fig.~\ref{fig:qhd} where the parameters used are the same as in Ref.~\cite{Lim:1989ma}. It is striking how much the transverse instabilities differ between the two models. What these results tell us is that including the medium modification of the quark structure of the nucleon pushes the transverse instabilities to much larger densities, larger than the applicability of the model. In other words, they are erased completely.

\subsection{Moat Regimes}

The most interesting finding, however, concerns the moat regimes. Although the instabilities towards inhomogeneous phases are washed away by the scalar polarizability, the moat regimes are all over the place. In both plots in Fig.~\ref{fig:qmc_sym} we show in light gray the \textit{global minimum} of the SDF where there is no zero crossing. For lower densities, around the instabilities related to the liquid-gas transition, the SDF is monotonic. After the instabilities go away, just after saturation density, a nonzero global minimum of the SDF takes place, which characterizes a moat regime. Of course, when there is an instability of $\epsilon_T$ we no longer have to show the global minima because, since the transverse part of the dielectric function shows up squared in the SDF, there will be two degenerate zeros on top of the blue lines in Fig.~\ref{fig:qmc_sym}.
In Fig.~\ref{fig:qhd} we also show the minimum of the SDF for the QHD model in light gray (where there is no zero cross of $\epsilon_L$). Here also, the moat regime takes place immediately after the liquid-gas transition. However, in the QHD model, there are instabilities all over the place.
In a way, this difference between the two models is consistent with the findings of several other references (e.g. \cite{Buballa:2020nsi,Pannullo:2022eqh,Pannullo:2024sov}) where it is shown that, for NJL like-models in the context of quark matter, the inhomogeneous phases are strongly dependent on the parameters, but the moat regimes are robust. 

\begin{figure}[b]
    \centering
    \includegraphics[width=\linewidth]{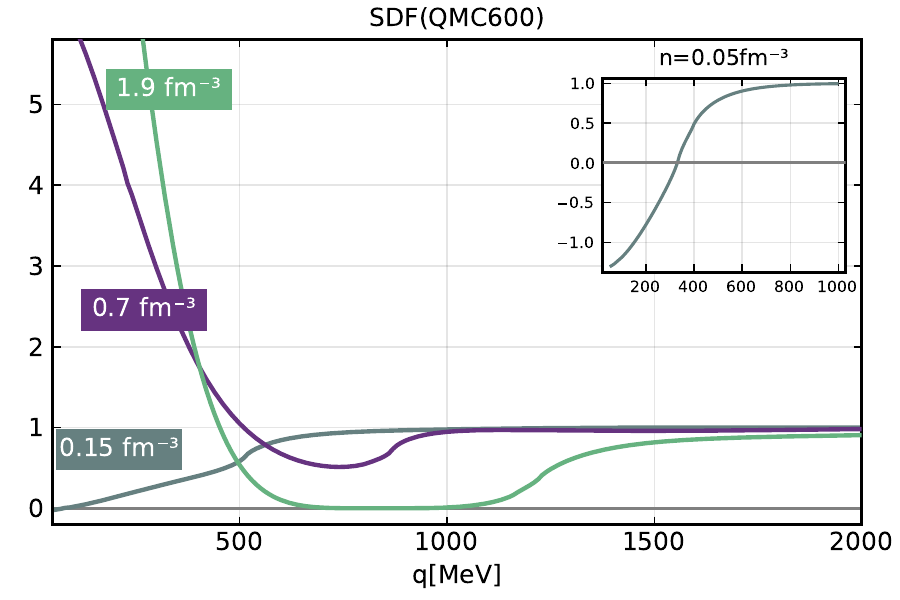}
    \caption{Static dielectric function for specific densities as a function of the three momentum. For $n=0.15$~fm$^{-3}$ we see that there is an instability for zero momentum, i.e., a homogeneous instability. This is characteristic of the liquid-gas coexistence region, the inset shows a lower density instability  of the same nature. For $n=0.7$~fm$^{-3}$ it doesn't cross zero, however, it also clearly show a non-zero minimum which characterises a moat regime. For $n=1.9$~fm$^{-3}$ we see an inhomogeneous instability.}
    \label{fig:moatqmc}
\end{figure}

In Fig.~\ref{fig:moatqmc} we show the full SDF as a function of the momentum for fixed number densities. It is clear that even for densities where the SDF does not touch the zero line, its behavior is clearly moat-like. The global minimum of a moat-like static correlation function in momentum space corresponds to the effective wave vector of the oscillatory behavior of this correlation function in configuration space. The fact that they are robust between the two parameterizations of the QMC model and also numerically very close to the QHD result for densities where we do not find instabilities shows that the scale of this wave-vector is quite robust.

It is clear then that the moat regime is a very present feature of symmetric nuclear matter for a wide range of densities.

\subsection{Pure Neutron Matter}
The same calculation can be performed for pure neutron matter (ignoring the $\omega-\rho$ mixings for simplicity). We see that all high-density instabilities vanish, but the moat regime is still there, which is also consistent with previous literature. In Ref.~\cite{Pannullo:2021edr}, the isospin-assymetric case was investigated in an NJL model, which shows inhomogeneous phases of quark matter in the isospin-symmetric case. There also, it was observed that the inhomogeneous instabilities shrink for larger isospin asymmetry.
 
\section{Summary and Conclusions}
Within the framework of the QMC model, we perform a stability analysis of homogeneous nuclear matter. We see that inhomogeneous instabilities are completely washed out in the model. If we compare this with the classical RMF result in Fig.~\ref{fig:qhd}, it shows a completely different story. Although it is interesting to see this, the far more interesting conclusion concerns the moat regimes. For both the QMC model (with a larger and lower scalar channel mass) and the classical RMF model, the moat regime starts at roughly the same density, and, even more remarkably, the wave number that marks the minimum of the static dielectric function seems to be roughly the same in every calculation, at least until a density of $0.41$~fm$^{-3}$, which is already way too high anyway. Although we do not show explicit plots for the sake of brevity, we have also performed this calculation for pure neutron matter, and there also the moat regime appears. This is indeed a very general feature of these models, and we can probably expect them to play a role in dense nuclear physics across the board.

In multiple works in the literature, e.g. Refs.~\cite{Pisarski:2020dnx,Lee:2015bva,Hidaka:2015xza}, it has been remarked that inhomogeneous phases might be disordered by fluctuations of Nambu-Goldstone bosons. When they are disordered, what is left is no longer crystalline, but still in a moat regime, either a liquid crystal \cite{Hidaka:2015xza} or the so-called quantum pion-liquid \cite{Pisarski:2020dnx}. These works, however, all consider quark matter. In the context of nuclear matter, recently, in Ref.~\cite{Kochankovski:2025thf} the authors considered the consequences of hyperons in a moat regime for compact stars. Interestingly enough, the possibility of nuclear matter in a liquid-crystal phase is not an exactly new topic of research; it was already put forward in the 90's \cite{Migdal:1990av}. In this work, we have shown that this is far from unrealistic and should be considered likely to happen in dense nuclear matter.

\section*{Acknowledgements}
The authors would like to thank Anthony W. Thomas and Fabian Rennecke for discussions and comments about the contents of this manuscript.

This work was partially financed by Funda\c{c}\~ao de Amparo \`a Pesquisa do Estado de S\~ao Paulo (FAPESP), grant nos. 2024/13426-0 (TFM) and 2018/25225-9 (GK), and Conselho Nacional de Desenvolvimento Cient{\'i}fico e Tecnol{\'o}gico (CNPq), grant no. 309262/2019-4 (GK).

\

The data that support the findings of this article are openly available \cite{Motta:2025xop}

\bibliographystyle{ieeetr} 
\bibliography{apssamp}

\end{document}